\documentclass[runningheads]{llncs}
\usepackage{graphicx}
\usepackage{enumitem}
\usepackage{color, colortbl}
\definecolor{Gray}{gray}{0.9}
\definecolor{MyBlue}{rgb}{0.0,0.404,.8}
\definecolor{MyBlueBorder}{rgb}{0.027,0.459,0.91}
\usepackage{multicol}
\usepackage{stfloats}
\usepackage{mathtools}
\usepackage{amsfonts}
\usepackage{listings, xcolor}
\usepackage{caption}
\usepackage{subcaption}

\begin{document}
\title{Blockchain for IoT: A Critical Analysis Concerning Performance and Scalability \thanks{Supported by RMIT RTS Program.}}
\titlerunning{Blockchain for IoT: A Critical Analysis}

\author{Ziaur Rahman \and
Xun Yi \and
Ibrahim Khalil \and 
Andrei Kelarev}
\authorrunning{R. Ziaur et al.}
%
\institute{RMIT University, Melbourne VIC 3000, Australia \\
\email{\{rahman.ziaur, xun.yi, ibrahim.khalil, andrei.kelarev\}@rmit.edu.au}}
\maketitle              
\begin{abstract}
The world has been experiencing a mind-blowing expansion of blockchain technology since it was first introduced as an emerging means of cryptocurrency called bitcoin. Currently, it has been regarded as a pervasive frame of reference across almost all research domains, ranging from virtual cash to agriculture or even supply-chain to the Internet of Things. The ability to have a self-administering register with legitimate immutability makes blockchain appealing for the Internet of Things (IoT). As billions of IoT devices  are now online in distributed fashion, the huge challenges and questions require to addressed in pursuit of urgently needed solutions. The present paper has been motivated by the aim of facilitating such efforts. The contribution of this work is to figure out those trade-offs the IoT ecosystem usually encounters because of the wrong choice of blockchain technology.  Unlike a survey or review, the critical findings of this paper target sorting out specific security challenges of blockchain-IoT Infrastructure. The contribution includes how to direct developers and researchers in this domain to pick out the unblemished combinations of Blockchain enabled IoT applications. In addition, the paper promises to bring a deep insight on Ethereum, Hyperledger blockchain and IOTA technology to show their limitations and prospects in terms of performance and scalability.

\end{abstract}

\section{Introduction}

Blockchain and IoT have been able to show immense effectiveness and potential for future improvements to productivity when being applied in collaboration. Therefore, how they could be employed to install end-to-end secure and sensor embedded automated solutions has become a frequently asked question. The world has already been surprised to experience the beautiful adaptations of different IoT solutions, ranging from healthcare-warehousing to transportation-logistics \cite{griggs2018healthcare}. Existing centralized Edge and Fog based IoT infrastructure may not be that scalable, secure and efficient to mitigate broader enterprise challenges. Mostly, emerging IoT solutions concern network of sensor-enabled smart appliances where it facilitates the services on the cloud of physical devices varies from modern car to smart-home utensils. In essence, an immutable timestamp ledger used for distributed data including either payment, contract, personal sharing and storing or supply chain and health care expected to impact several sectors due to its salient features such as immutability, distributed structure, consensus-driven behavior and transparency \cite{nakamoto2008bitcoin}. 

\subsection{Blockchain's Potentials to be peer with IoT}

There are several reasons why blockchain can be very promising to ensure efficiency, scalability and security of the IoT arrangement. Firstly, it has a proven cryptographic signing capability to perform end-to-end cryptographic message transfer. It is able to enable asset functionality to provide good governess \cite{GSMA2018}\cite{ferrag2018blockchain}. Moreover, it can address custodial tracking of asset transmission in a global logistic phenomenon. Besides, next we give a list of several emerging issues and their corresponding blockchain potentials as follows. 

\begin{enumerate}[label=\roman*]
   \item Free global infrastructure that offers 
   \begin{itemize}[label=\textendash]
       \item Blockchain is leveraging and reliable 
   \end{itemize}
   \item Data belonging in the Edge-Network the final destination to the appliances
   \begin{itemize}[label=\textendash]
       \item Blockchain has the ability to be replicated rapidly and consistently to every closest node for which it will certainly will be cost-effective
   \end{itemize}
   \item Hack-proof cryptography eliminating attacks
   \begin{itemize}[label=\textendash]
       \item Already proven to be resilient to the popular attacks - for example: if you are looking to protect a power grid or if you are looking to protect high value asset it becomes natural to use blockchain 
   \end{itemize}
   \item Record proof of life for industrial assets in an irreversible ledger
   \begin{itemize}[label=\textendash]
       \item people can even associate it to an asset to verify its validity which might be increasing the revenue of the company as there is no counterfeiting. 
   \end{itemize}
   \item Track-chain of custody assets on Transportation or sale
   \begin{itemize}[label=\textendash]
       \item we not only can verify but also can track when those are sold to a different individuals allowing us to gain the types of metrics that we would necessarily lose thus we can provide insight to companies. 
       \item Full redundancy providing a hundred percent uptime and assuring message delivery 
   \end{itemize}
  \end{enumerate}
  
The contribution of this work is to figure out those trade-offs the IoT ecosystem usually encounters because of the wrong choice of blockchain technology.  Unlike a survey or review, the critical findings of this paper target sorting out specific performance and scalability challenges of blockchain-IoT Infrastructure. The contribution includes how to direct developers and researchers in this domain to pick out the unblemished combinations of Blockchain enabled IoT applications. The claimed contributions are justified through the respective sections of the paper. The Section 3 of this paper discusses Blockchain suitability to eliminate the problems that emerges because of Blockchain and IoT integration \cite{access}. The later sections explains how existing solution namely Microsoft Azure adopts different Blockchain platforms such as Ethereum, Hyperledger, etc.  The following section illustrates Blockchain potentials for specific IoT issues.  The challenges come to light while a sensor-enabled system finds its devices, managing access control, etc through respective use-case anaysis. Furthermore, the analysis justifies the smart contract compliance for IoT system along with data integrity and confidentiality loop-holes.

Therefore, the article is organized as follows: the preceding  section and the introduction throughout its subsections talk about why blockchain is necessarily applicable in the Internet of Things (IoT) \cite{xuyang}. Section~2 is a bit about blockchain internal design and its tailored categories leading with part 3 where an OSI like blockchain open system structure redrawn following some previous works. Sections~4 5 and 6 portray the comparative analysis with contemporary technologies including Hyperledger, IOTA and Microsoft Azure IoT architecture. Then the following section summaries with a brief table and graphs showing the challenges and proposed solutions at a glance as well as its applicability concerning the throughput and latency. A set of use cases where blockchain is an inevitable peer of IoT mentioned before the conclusion on top of advantages \& application.

\section{Blockchain Preliminaries}

The blockchain is a means of removing the need of traditionally created trust created through intermediaries in the distributed systems. A blockchain enables trusts among untrusting entities within a common interest. Thus, it helps to form a permanent and transparent record of exchange of processing, ignoring the need for an intermediary. The terms blockchain and distributed ledger often used interchangeably, but they are not always the same thing. Blockchain is about the exchange of value instant, decentralized, pseudonymous value transfer which is now possible. It can ensure ledger building by preserving a set of transactions shared to all participating users, where the new one is necessarily verified and validated by others (\cite{salman2018security}, \cite{fernandez2018review}). Adjoining brand-new transaction usually called mining which demands solving the complex and substantial computational puzzle which in nature is a complicated answer but simplest to authenticate using a chosen consensus mechanism in the network of untrusted and anonymous nodes. That indeed has brought enormous transparency for a BC-enabled applications. Significant resource constraints required to facilitate the consensus algorithm by which it restricts unauthorised blocks from joining the network. Besides, communication among nodes are encrypted by changeable public keys (PK) to prevent tracking itself, thus it has been able to  draw attention in non-monetary application \cite{ferrag2018blockchain}, \cite{dinh2018untangling}.

A sample chain of blocks can be delineated where each block depicts the hash of the previous block, time stamp, transaction root and nonce created by the miner~\cite{minoli2018blockchain}.

\begin{figure*}[h!]
\centering\includegraphics[width=\textwidth]{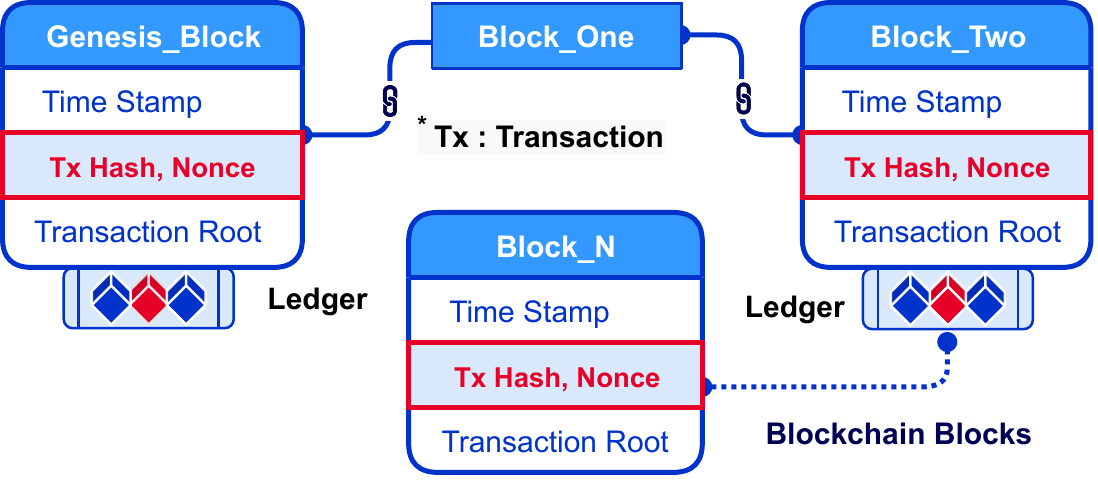}
\caption{At a glance view of chain of different blocks with nones, time stamp, transaction data and hashes}
\end{figure*}

It has already been able to show its potentials in particular in this field by super setting secure smart device authentication to ensure uncompromised communication, decentralized data formulation or even automatic data purchasing and others. Thus, we can conceivably estimate that an emerging phenomenon of IoT utensils would be equipped with the Internet to ease every aspect of human life\cite{hammi2018bubbles}.

\begin{figure*}[ht!] 
\centering\includegraphics[width=4.1in]{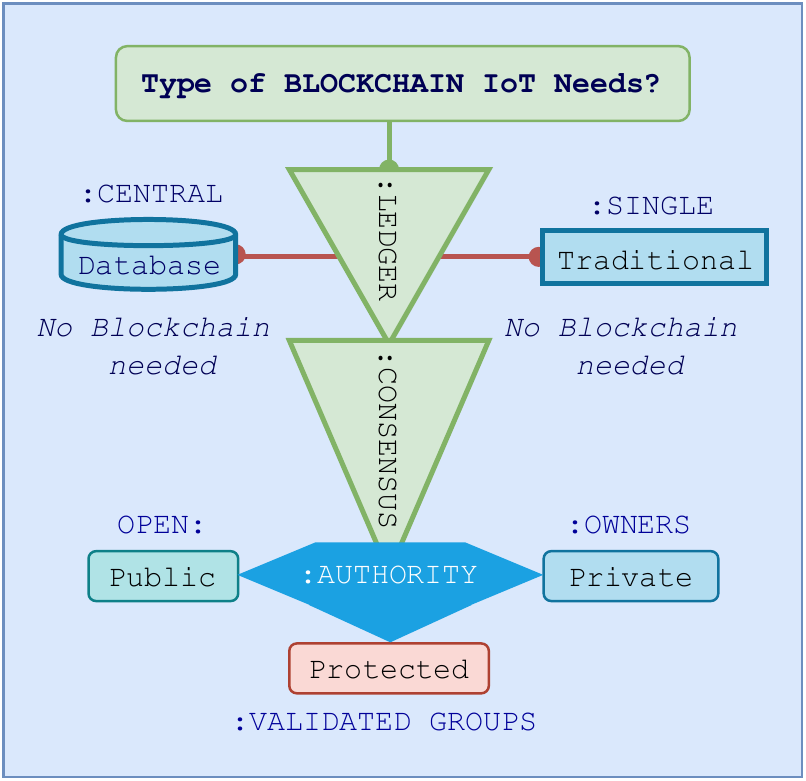}
\caption{Blockchain Grouping according to the requirement analysis. It shows either the need of the particular type of blockchain or the usual approach is able to meet the demand}
\end{figure*}

\subsection{Blockchain Category}

The following Figure 2 shows the comparative classification of blockchain ledger concerning the considered ledger accessibility.

\textbf{Public Ledger based Blockchain}. It is also often known as permissionless blockchain as anyone can send, verify and read transactions on the network even able to get and run the codes on their machine to take part in mining process using consensus \cite{dinh2018untangling} algorithms. It has the maximum anonymity and transparency even though any user unidentified are allowed to send, read and validate the incognito transaction. Ethereum and Bitcoin are the typical examples of public blockchain. \textbf{Private Ledger based  Blockchain.}It controls the access privileges by restricting read and modification right to a particular organization; thus, it does not require a consensus mechanism or mining to ensure anonymity. In some instances, the read authority kept restricted to an arbitrary level, but mostly the transaction editing is strictly permissioned. The ledger-building process for coin supervised by Eris and Monax or the Multichain could be said to have private-typed blockchain techniques \cite{fernandez2018review}. It dserve mentioning that Ethereum now has permissioned Blockchain, such as Quorum. \textbf{Protected Ledger based Blockchain.}
Protected Blockchain is also known as Consortium/federated \cite{ziacomsoc} or and in some cases it is called hybrid or publicly permissioned blockchain which is maintained within the authority of a group of owner or users \cite{saraf2018blockchain}. Hyperledger by Linux Foundation and IBM \cite{griggs2018healthcare}, Services of R3 with Corda or Energy Web Foundation are example of protected type of Blockchain \cite{abdella2018peer}. However, the required blockchain type shown through Figure 5 based after the ledger, consensus and the dependability of the type of authority. Figure 2 shows that if the system has a centralized or single ledger system, no category of the blockchain is needed there. However, if the authority is restricted within a validated group, then protected one seems to suit more than public or private ledger based blockchain system \cite{kshetri2017can}. Beside, above types of Blockchain, this manuscript explains performance comparison of IOTA. The founders of IOTA have described its ledger as a public permission-less backbone for the Internet of Things that enables interoperability between multiple devices \cite{ziacomsoc}. That means, it will enable transactions (tx) between connected devices, and anyone on the network can access its ledger.

\begin{table}[h!] 
\caption{Comparison among Different  Popularly used Consensus Mechanisms}
\begin{center}
\begin{tabular}{l|c|c|c|c} \hline
\multicolumn{1}{c }{\textbf{Attributes}}
& \multicolumn{1}{c}{\textbf{PoW}}
& \multicolumn{1}{c}{\textbf{PoS}}
& \multicolumn{1}{c}{\textbf{BFT}}
& \multicolumn{1}{c}{\textbf{PoA}} 
 \\ \hline 
Category      &  Public  &  Pub/Protected  & Private  & Protected \\ 
Random      &  No  &  Yes  &  No &  No \\ 
Throughput      &  Little  &  Big  & Big  &  Big \\
Token      &  Has  &  Has  &  Not & Native  \\ 
 P-Cost & Has &  Has  &  Not & Not  \\ 
Scalability & Big  &  Big  &  Little & Medium  \\ 
Trust & Trustless  & Trustless  &  Semi & Trusted  \\ 
Reward & Yes  &  No  &  No & No  \\ 
Example & Bitcoin  &  Ethereum  &  Hypeledger & Kovan  \\  \hline
\end{tabular}
\label{}
\end{center}
\end{table}

\section{Blockchain Suitability for IoT}

As explained earlier, Blockchain can solve all IoT issues. There are several problems where a centralized database can be a good solution instead of applying blockchain. The following section illustartes Blockchain applicability for IoT and the attributes that necessarily need to be discussed before applying it any IoT use cases. 

\subsection{Comparison among Consensus Protocols}

Table 1 shows the comparison among different consensus mechanism. It illustrates that Proof of Work (PoW) or Proof of Stake (PoS) require significant computational resource, whereas Byzantine Fault Tolerance (BFT) and Proof of Authority (PoA) have higher throughput in comparison to its peers. But in either case of BFT and PoA scalability can be a challenge. Another thing is that they have a token in dependability, which seems to be working fine for IoT nodes. In case of scalability and Overheads, blocks are broadcast to and verified by all nodes with a quadratic increment of the traffic, and intractable processing overheads that demand huge extensibility whereas IoT devices (e.g., LORA) have limited Bandwidth connections \cite{hammi2018bubbles}. For delay/latency: IoT devices have stricter delay requirement (e.g., Smart Home Sensor unlikely to wait) whereas BTC can take approximately 30 minutes to confirm a transaction. Even it has security overheads as it has to protect the double spending seems inapplicable for IoT. For bitcoin, the throughput is 7/Transaction which would go beyond such limit due to huge interaction among nodes in IoT. Therefore, after bitcoin many have been opted for BFT based Hyperledger (\cite{griggs2018healthcare}, \cite{dinh2018untangling}) or non-consensus driven approach such as IOTA \cite{popov2017equilibria}. The applicability of different blockchain platforms based on those consensus protocols and non-consensus approach discussed with this \cite{dorri2017towards}.

\begin{figure*}[ht!]
\centering\includegraphics[width=\textwidth]{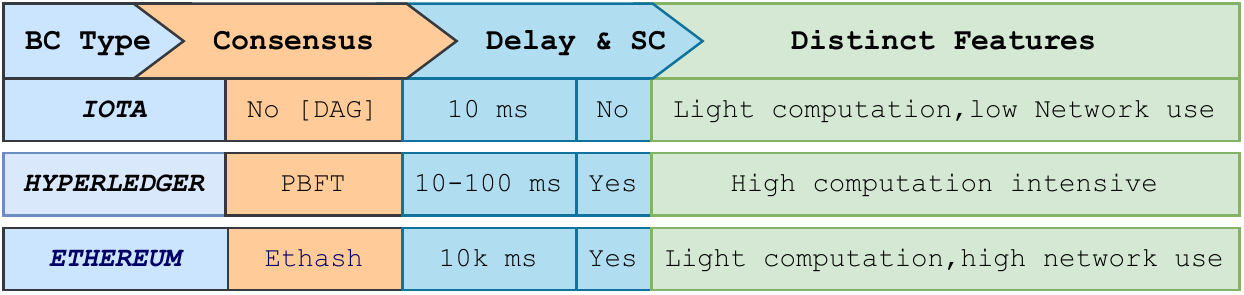}
\caption{The both part of the above figure shows the comparative Analysis among Ethereum \cite{buterin2014next}, Hyperledger \cite{griggs2018healthcare} and IoTA \cite{popov2017equilibria} and before applying those as a mean of IoT Performance and Scalability}
\end{figure*}

\subsection{Etherem, Hyperledger and IOTA}

\textbf{Ethereum} emerged intending to compete with bitcoin is a flexible blockchain platform with required smart contracts and proof-of-work consensus mechanism called Ethash. This associates with Directed Acyclic Graphs (DAG) \cite{popov2017equilibria} to generate the probabilistic hash. It ensures robust extensibility for the IoT applications, including some efficiency trade-offs. As Ethash works upon PoW, Ethereum requires around 20s to append a new block after mining \cite{ferrag2018blockchain} \cite{dinh2018untangling}. Secondly, \textbf{Hyperledger} is a permissioned and protected type of blockchain. It commonly applies access control, along with chaincode-based smart contracts and consensus with existing Practical Byzantine Fault Tolerance (PBFT) \cite{aitzhan2018security} \cite{dinh2018untangling}. It includes anchors of trust to base certificate-authorities as an increment to the asymmetric cryptographic approach and digital signature properties with SHA3 or ECDSA. Hence, its smart contracts implementation involves the chaincode that has a self-execution ability such as asset or resource transferring among network-peers in huge time. This latency is low among comparative distributed ledger implementations. Fabric has been chosen as blockchain medium by IBM in their Bluemix-Watson IoT architecture, which has been shown by the respective section hereafter. \textbf{IOTA} which a unique distributed ledger in that it does not utilize an explicit blockchain at all; instead, it implements a directed acyclic graph of transactions – instead of blocks of multiple transactions that link together, each transaction approves and links back to two other transactions. IOTA Tangles has immense potentials to be efficiently adapted with IoT to ensure security and privacy by ensuring maximum throughput. Fig. 3 shows the comparative analysis among Ethereum, Hyperledger and IOTA in terms of performance and scalability.

\subsection{Azure IoT Workbench}

Figure 4 illustrates the Azure IoT workbench that facilitates client-side application for both mobile and web system depending on the smart-contract. It purposes to verify, retrieve and test applications or entertain new use-cases there. It brings a user-interface to interact with the end-user for appropriate tasks. Besides, entitled individuals are given to permission accessing the administrative console with different functionalities such as uploading and deploying smart contracts depending on certain roles. 
As depicted in the figure, the workbench has a gateway-service API standing on the representational state transfer (REST) API reproduces and delivers messages to event-broker while attempting to append data to blockchain. Queries are submitted to off-chain-database when data is requested. The database mostly the SQL contains a replication of all chained meta-data and bulk data that issues relevant configuring context for the smart contracts supported. Thus, the users with developer role are allowed to get accessed the gateway servicing API to develop blockchain apps without depending on the client/end-user solutions. In case of message breaking for incoming data, users who desire to circulate messages thoroughly to the Azure workbench can submit data directly to the service bus there. For illustration, this API solutions for system integrated confederation or sensor based tools. Apart from this events are held during the life-time of the application. It can be caused by the gateway API or even inside ledger and its alerting trigger downstream-code based on the event so far occurred.  Microsoft Azure consortium usually able to locate two different kinds of event consumers. First one gets activated by the events lies on the blockchain to manipulate the off chain SQL storage while the rest responds capturing meta-data for the events brought by document upload and storage related API. Figure 4 shows how Microsoft (MS) Azure IoT work bench adapts different Blockchain farmeworks. It also portrays that MS Azure architecture can facilitate Hyperledger Fabric (HLF), CORDA R3, or IOTA. The IoT Hub is connected to the IoT sensors and its bus is enjoined to the Transaction Builder. MS Azure  can be proof that how exisiting IoT workbench can be implemented for the scalable and secure IoT service.


\begin{figure*}[ht!] 
\centering\includegraphics[width=\textwidth]{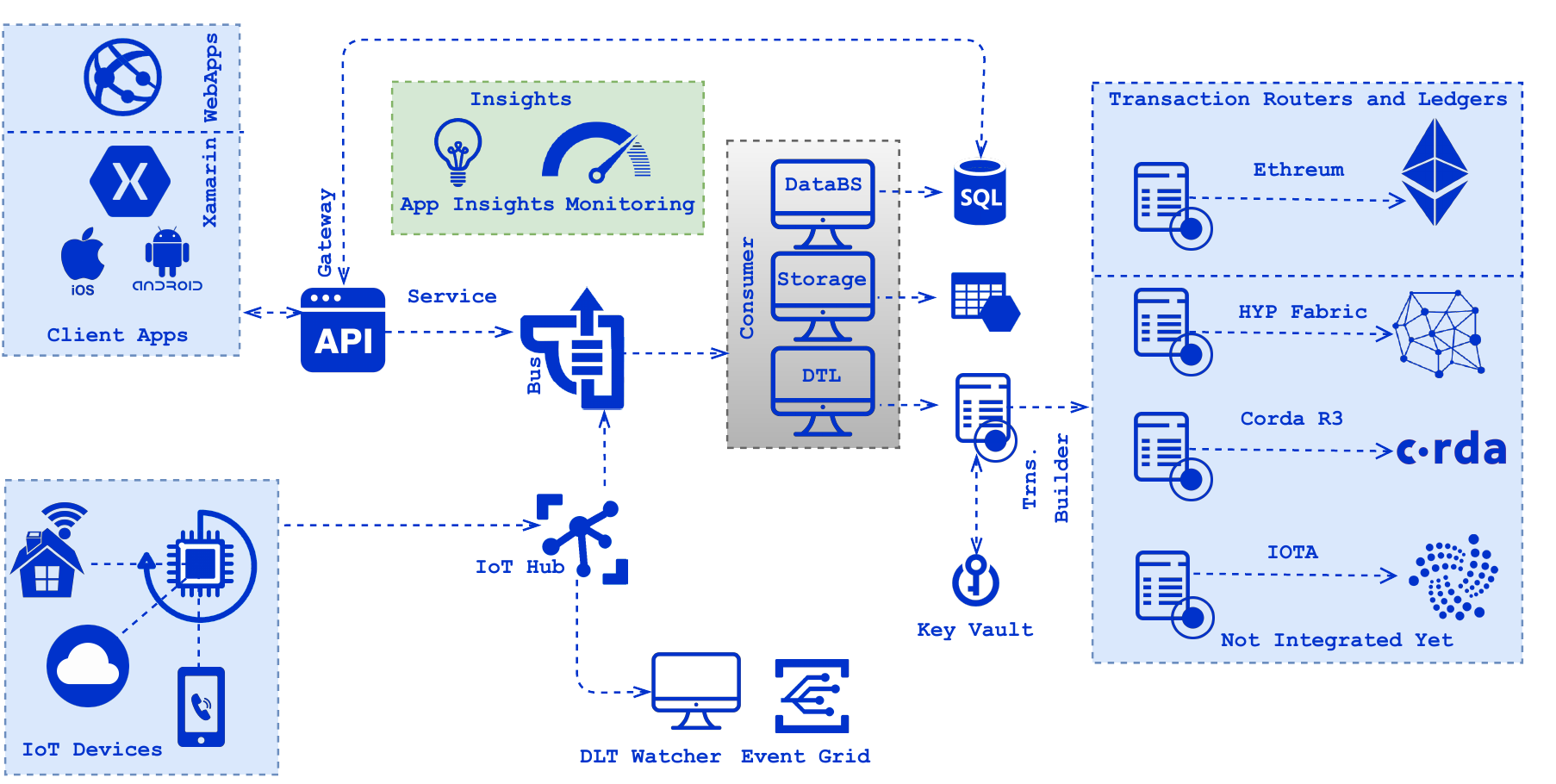}
\caption{Microsoft Azure Blockchain Architecture that has been integrated with Ethreum for securing IoT appliances. The COrda, Hyperledger and IOTA could be incorporated just like Ethreum as said by Azure}
\end{figure*}

\begin{figure*}[ht!] 
\centering\includegraphics[width=\textwidth]{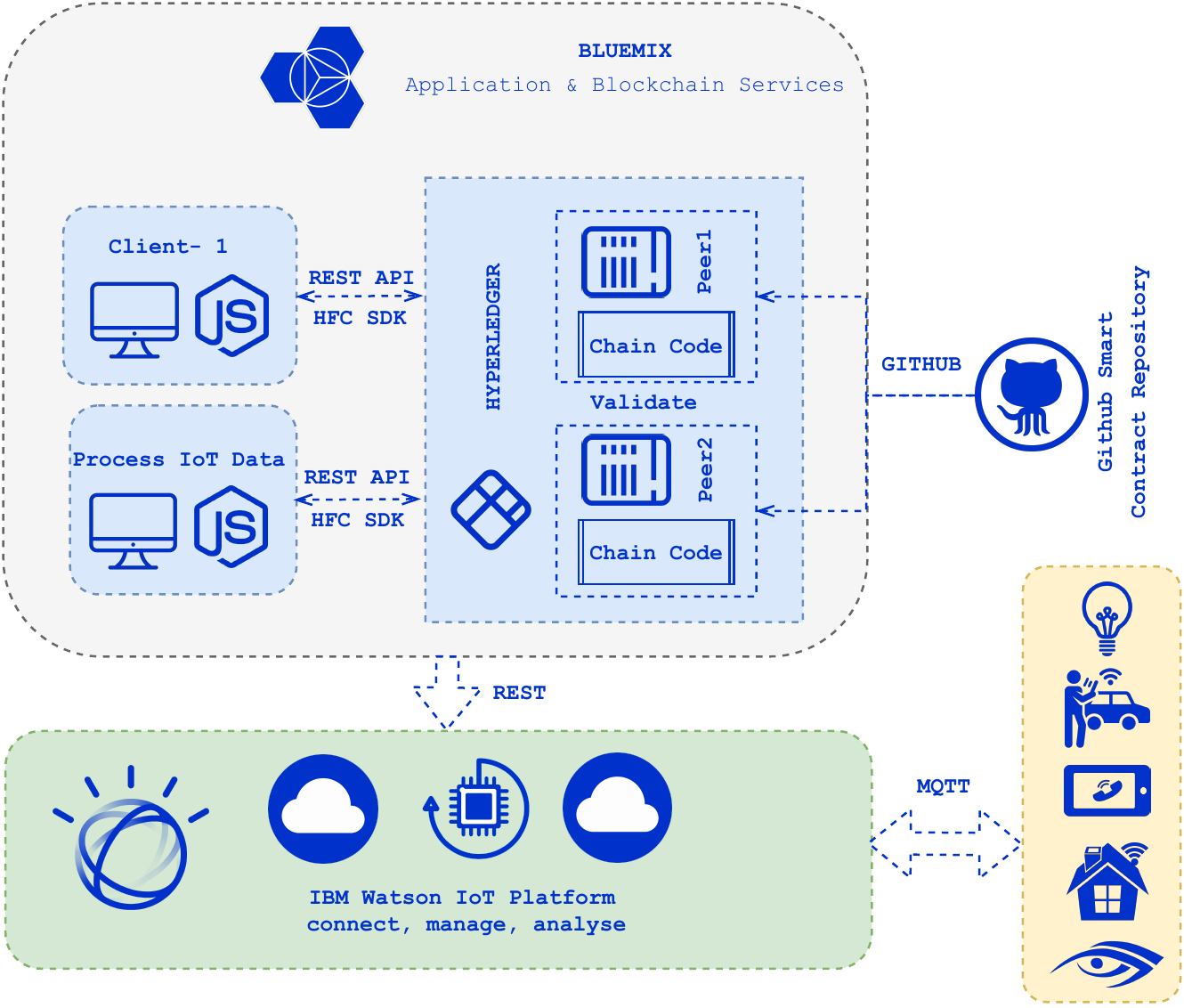}
\caption{IBM Watson and Bluemix Integration for IoT-Blockchain Service. Watson communicates the IoT devices, where Bluemix facilitates Blockchain network on top of smart contarct repository from Github}
\end{figure*}

\subsection{IBM Blockchain Integrated IoT Architecture}

The IBM Blockchain architecture for IoT solutions comprises three principal tiers; each has customized roles in its own side. The Figure 5 depicts the high level IoT architecture incorporating Hyperledger Fabric as Blockchain service, Watson as IoT Platform and Bluemix as cloud environment \cite{griggs2018healthcare} \cite{eckhoff2017privacy}. The IBM IoT architecture can be divided into several components as shown by Figure 5. It has been discussed with its three tiers, service execution process and along with the challenges it encounters. It also shows the IBM Blumix IoT working procedure. During execution, data collected from smart devices and intelligent sensors are dispatched to Watson using ISO standard (Message Queuing Telemetry Transport (MQTT) protocol. Specific blockchain proxy works to send data from Watson to the chain-code of the Hyperledger Fabric depending on the settlement. Hence, the transactions get executed in the Cloud. The solution-components associated in the execution process have been enlisted as below:

\subsubsection{Smart Contarct as Chaincode}
Instead of using Bitcoin or Ethereum like smart contract Hypeledger fabric adapts the chaincode written with Go. It shapes the core distributed ledger solutions and necessarily epitomize the desired business logic. Each transaction call out there is carefully preserved and prevailed as expected blockchain transaction. As Fabric contract is the chaincode, it requires implementation with the certain APIs so, the chaincode needs to get registered with the services using those pre=defined APIs.  Hyperledger Fabric Client (HFC) Software Development Kits (SDK) ease developers to create Node.js applications able to maintain communication with the blockchain network. Here, the applications are registered and submitted transactions using APIs. IBM Blockchain aligned IoT Architecture on Bluemix offers several advantages such as trust, autonomy, scalability, security in the distributed network comprising multiple parties. Even though there are some challenges need to overcome. One of the significant issues is the power of computation, as IoT devices are usually low powered devices and have less computation capacity. Moreover, encrypting and transaction verification may require huge electricity. It can increase both energy consumption and expenses as well \cite{ferrag2018blockchain}. 

\section{Blockchain-IoT Challenges and Solutions}

Despite enormous engaging attributes of blockchain for IoT implementation, there are several challenges; each of which deserves proper concerning solutions before fruitful lodgement of Blockchain in the of IoT domain.

\subsection{Storage, Throughput and Latency Challenges}

Ethreum and Bitcoin have storage and Latency challenges as discussed earlier. The storage size has been increasing day by day as shown by Figure 6. It represents the incremental storage amount from 2015 to quarter August 2021. Blockchain platform that requires higher storage has lesser suitability for real-time system such as IoT. IoT system generates huge and voluminous data which indulges the chances of failure because of storage overhead. Figure 6 shows that in terms of storage the Ethereum seems more suitable than Bitcoin. Though only storage overhead is not the only standard to decide whether a particular Blockchain is suitable or not. But surely, it affects both the performance and scalability of the system.  Whereas, the following Figure 7 shows a comparison among Ethreum, Parity and Hyperledger with per second transaction amount labelled beside the bars. The data were found from \cite{dinh2018untangling}. They worked with blockbench collecting data from Yahoo Cloud Serving Benchmark (YCSB) a and Smallbank. It concludes by showing Hypeledger has the maximum throughput. While the second part shows Hyperledger works fine under 16 Nodes. The challenge is  here how it can be improved when more nodes will be incorporated as depicted by Fig. 7. Figure 7 shows the throughput Comparison among Ethreum, Ethereum (ETH) Parity and Hyperledger (HLF) fabric based on the Data found on using Blockbench. Though Parity is one of several implementations of Ethereum, it was considered as alternative Blockchain solution for IoT. Therefore, the Figure illustrates both the Ethereum (ETH) and ETH Parity. Hyperledger is a multi-project open source collaborative effort hosted by The Linux Foundation, created to advance cross-industry blockchain technologies \cite{ziacomsoc}. In this comparison we consider Hyperledger(HLF) Fabric only. The scalability challenges as figured out needs proper attention, otherwise the large-scale IoT system not that type of Blockchain after a certain volume of sensor integration.

\begin{figure*}[h!] 
\centering\includegraphics[width=\textwidth]{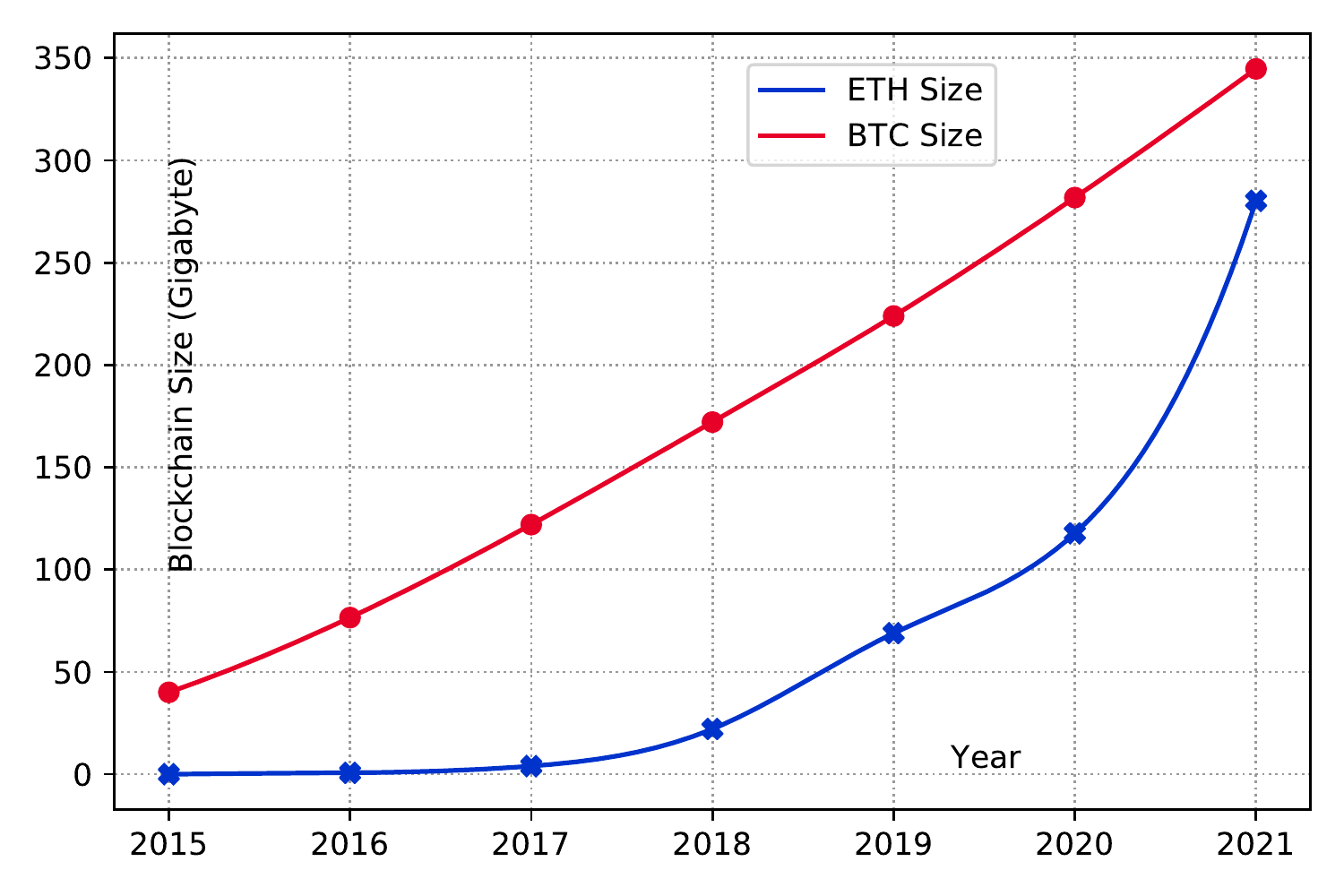}
\caption{Comparison of Chain-data storage between Bitcoin (BTC) and Ethreum (ETH) based on the online data provided by blockchain and statista and etherscan.io website}
\end{figure*}

\begin{figure*}[h!] 
\centering\includegraphics[width=\textwidth]{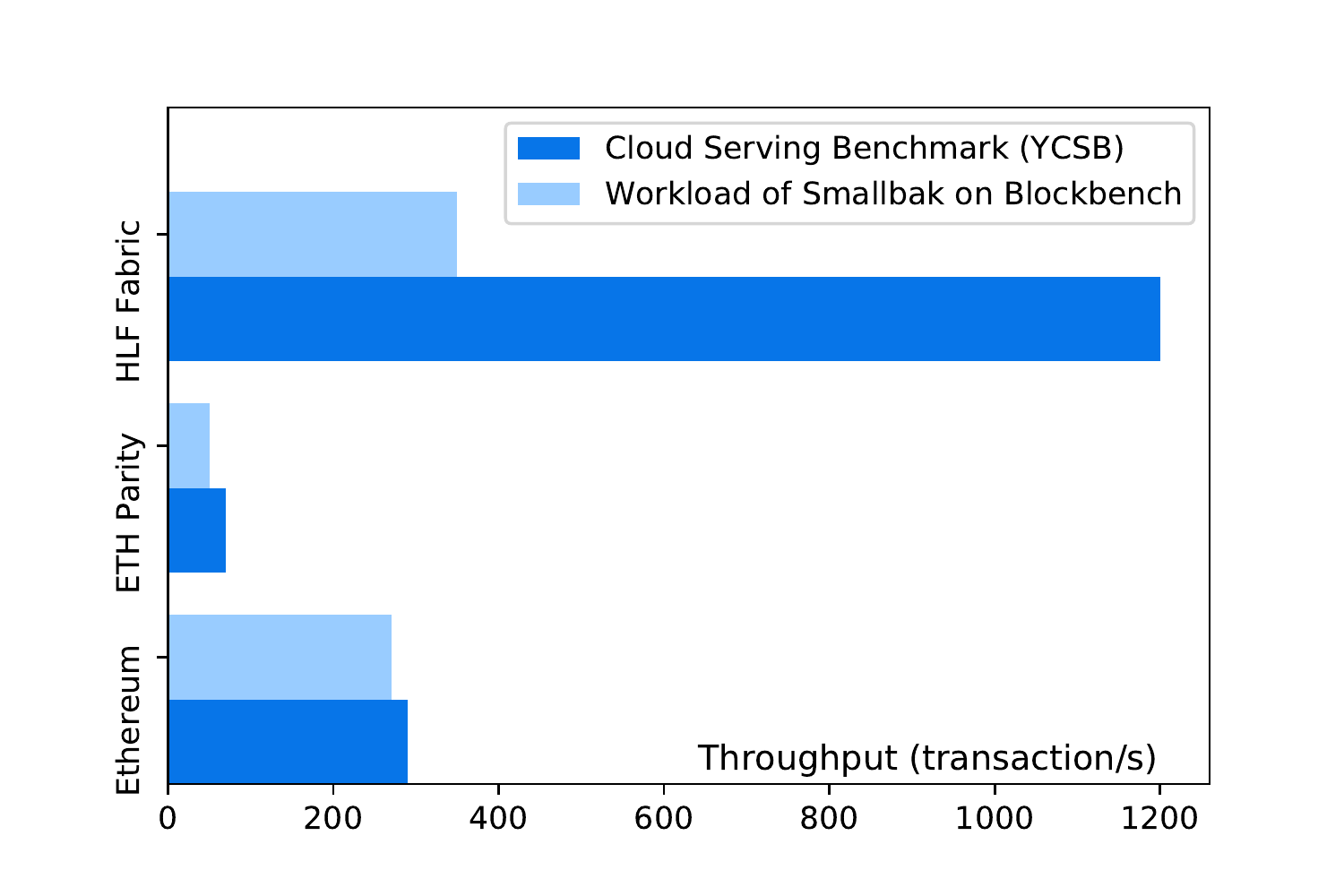}
\caption{Throughput (Performance Efficiency) comparison among Ethreum, Ethereum (ETH) Parity and Hyperledger (HLF) fabric based on Data in \cite{dinh2018untangling} using Blockbench framework.}
\end{figure*}

\subsection{Prominent Challenges and Solutions}

As IoT systems vary from smart coffee machines to complex automobiles, it is difficult to generalize all challenges in one table. The following Figure 9 describes some inevitable challenges and their possible solutions, respectively \cite{eckhoff2017privacy}. We have so far included seven possible challenges and respective blockchain solutions may require considerations before applying it for IoT architecture \cite{jiot}.



\section{Use case Analysis}

According to a review \cite{GSMA2018}, the research on blockchain and distributed ledger in association with the several mobile operators conducted by GSMA \cite{GSMA2018},  the emerging application of distributed ledger for blockchain can be put into three different sets ordering as Areas with common IoT Controls, Areas where IoT Appropriates and Areas with Particular IoT Solutions. The Figure 8 shows a relative study comparing among application areas with respect to three priority interests- maximum, medium and minimum. For Data Sharing as for illustration, three operators recommends it as medium while five of all suggests as the maximum priority of interest, leaving the access control application with minimum priority. As claimed by GSMA, data were collected with sincerely exploring all operators, still it deserves further before assessing for technical and industrial implementation. Considering conciseness and brevity, four use cases directly relating to performance, security and sclability that have been discussed in the following sections.


\begin{figure*}[h!] 
\centering\includegraphics[width=\textwidth]{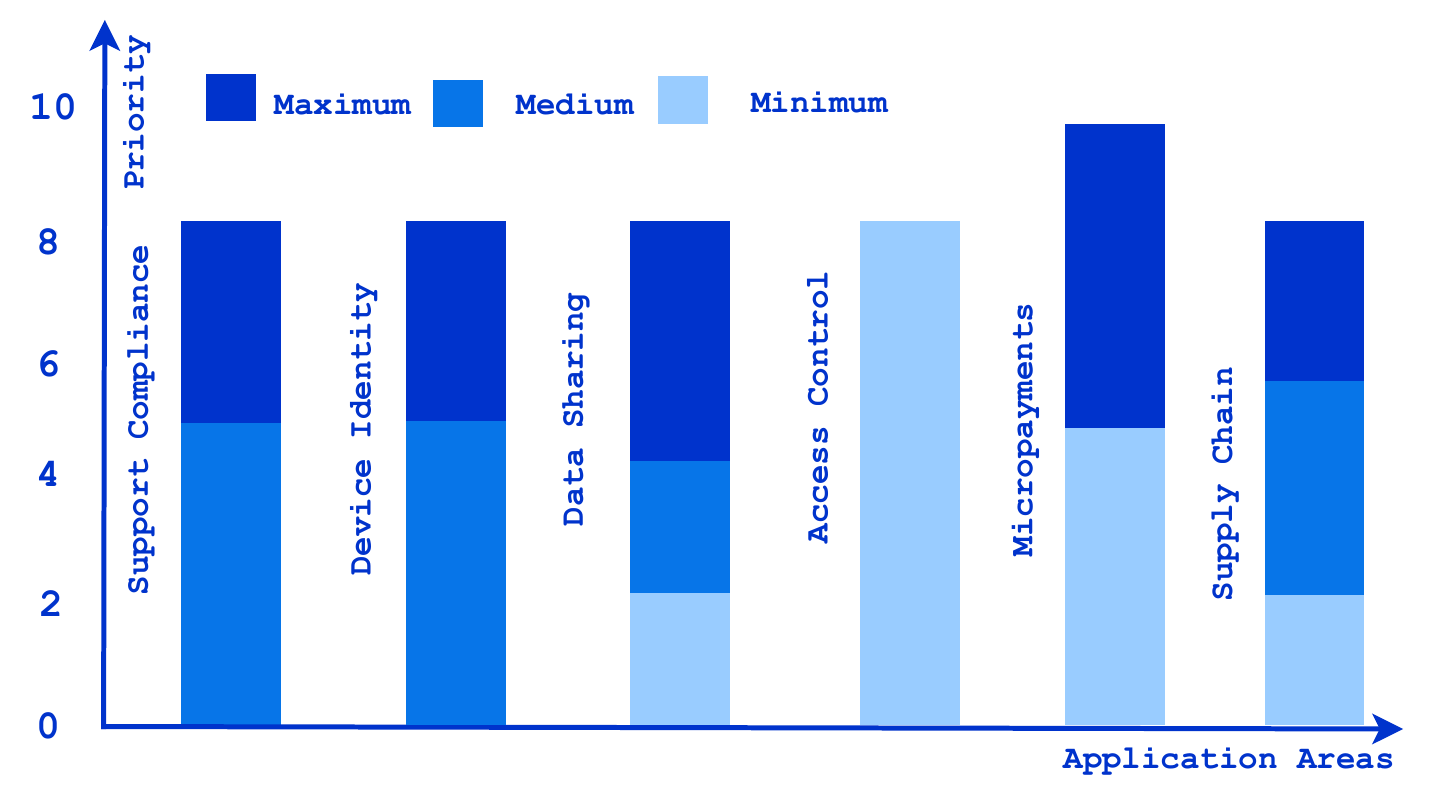}
\caption{Applicability of six considered Blockchain aligned IoT usecases according to the review by 10 operators plotting with their priority and applicability. The problems as illustrated through the usecases can be solved after incorporating Blockchain technology.}
\end{figure*}


\subsection{Use Case : Finding IoT Devices}
Retrieving and tracking identity information of the devices has been influencing factor with the growth of IoT enabling. The following cases will describe examples of finding intelligent devices in the IoT Network.   

\begin{itemize}[label=]
    \item \textbf{\textit{Case 1}}: Storing the original data and device status toward authentication. For example- For example identifying the manufacturing company or party if it has quality assurance accreditation including the life cycle status and validating the serial numbers provided.
    \newline
    \item \textbf{\textit{Case 2}}: Issuer signature verification according to the information stored in the ledger to make sure software updates from trusted sources.
    \newline
   \item \textbf{\textit{Case 3}}: Preserving and ownership device information such as hardware configuration, version information, boot code installation purposing to ensure privacy status check.
   \newline
   \end{itemize}
 
 


\subsection{Use Case : IoT Access Control}
A Monitoring and recording access control in inevitable in IoT network to preserve access control details for both physical and virtual resources. Therefore, the use-case for this can be as below-   

\begin{itemize}[label=]
    \item \textbf{\textit{Case 1}}: Virtual File sharing server uses the ledger to preserve identity of the individuals and application by securely assigning printing, saving or editing accessibility rights. For example – while a consumer made order for a good online is away from home, there is a risk in deliveries. So, customers can get benefit of using distributed ledger, instead of giving access to their home by using keys/address/codes prone to be misused.  
    \newline
    \item \textbf{\textit{Case 2}}: Issuer signature verification according to the information stored in the ledger to make sure software updates from trusted sources.
    \newline
\end{itemize}




\begin{figure*} 
\centering\includegraphics[width=\textwidth]{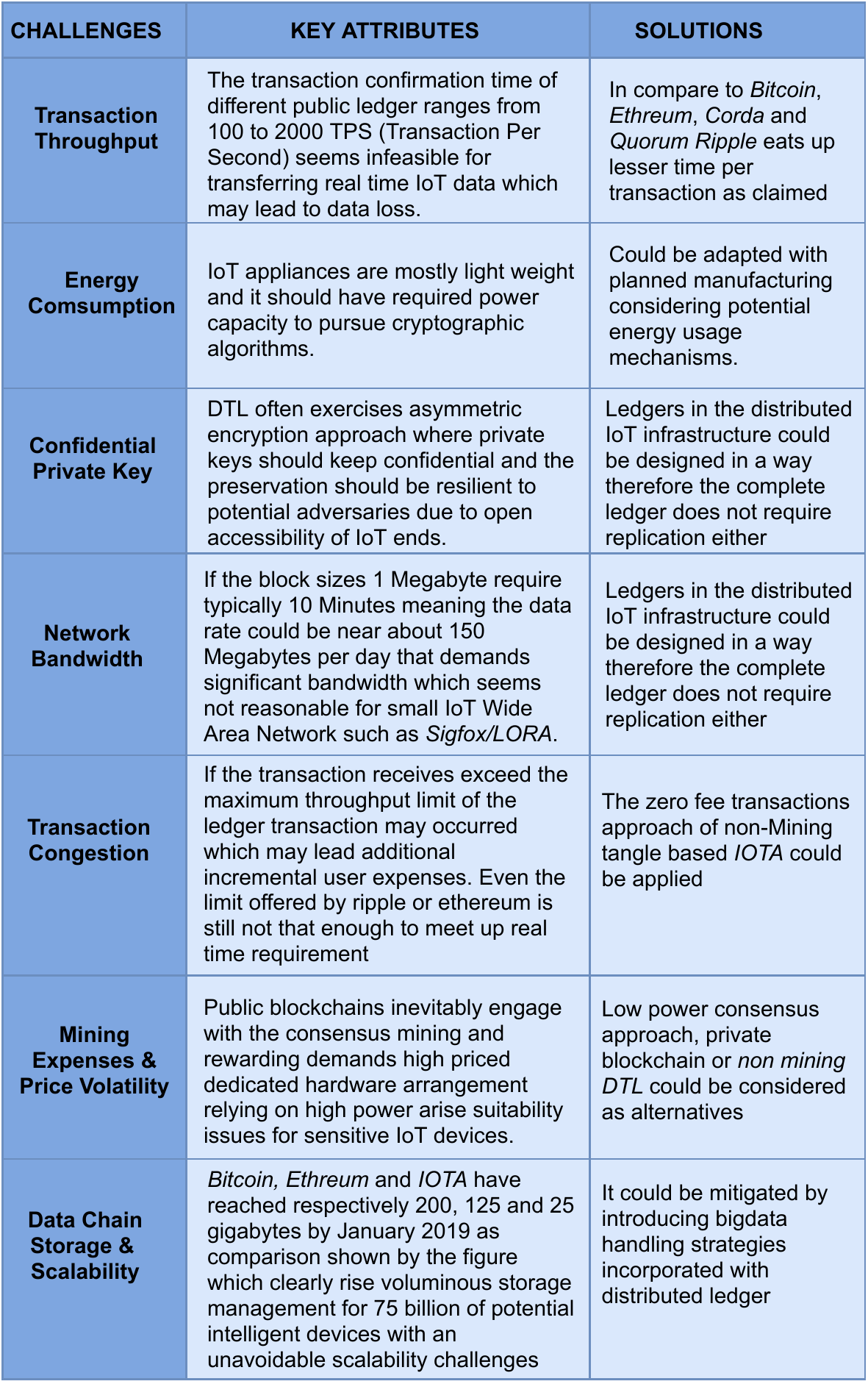}
\caption{Blockchain Implementation Challenges in the IoT and Probable Solutions Identified}
\end{figure*}

\subsection{Use Case : Supporting Smart Contract Compliance}
There are a lot of circumstances associating multiple companies where it is equally important to know whether the all of those are properly complied with. Thus, compliance is efficiently activated using blockchain smart contract. For example, let consider the following cases.  

\begin{itemize}[label=]
    \item \textbf{\textit{Case 1}}: Certain Individual sharing personal data with their health provider can control accessibility using distributed ledgers to make sure good data governance if it is only being accessed by the authorized medical professionals. In multiparty system, patient blood pressure should be only shared by pharmacy and general practitioner so that prescribed drugs can be easily dispensed.  
    \newline
    \item \textbf{\textit{Case 2}}: Suppose a person has to pay 2 Dollar extra for airport taxi pickup if the flight delays for 30 minutes. In a micro-insurance premium case like this lower cost feature of service delivery in the smart contract can automatically trigger it on arrival by determining whether that extra premium has been paid in order. Blockchain can play a vital role to address the issues mentioned in the use-cases. 
    \newline
    
    \item \textbf{\textit{Case 3}}: Driving particulars including license information, previous record of traffic rules violation, health and safety compliance either of a person or car need to get verified before one can drive a connected hire car. Even the car can upload journey information, servicing history, faults made by itself.  In a case where hundreds of thousand cars and drivers are affiliated with, smart contract and blockchain can easily provide required information within least hardship and delay. \newline
\end{itemize}




\subsection{Use Case: Data Integrity and Confidentiality}

It is often desired that the data sharing with keeping it confidential enough would be remarkably plausible in a distributed ledger framework \cite{dorri2017blockchain}. One of the significant features of blockchain is it can be applied to assert data integrity and IoT affiliated data effectively by maintaining the sequence of digital signatures and data hashes. A use case for this can as following \cite{surya}.  Selected challenges are outlined in the Fig. 9.

\begin{itemize}[label=]
    \item \textbf{\textit{Case 1}}: IoT Devices are expected to transfer information to the servers belongs to the manufacturing company. For example- intelligent thermostat connected with cloud services determines when to switch on and off depending of the current weather status can send data to the company about component wear. Existing solution such as Public Key Infrastructure (PKI) driven techniques can solve issues like this but Blockchain seems to more efficient deterring the need to reinvent processes with integrity and confidentiality.   
    \newline
    \item \textbf{\textit{Case 2}}: Home or office alarm machines can be controlled by different entity as per their access privileges assigned. In case of it is compromised by intruders it may require to remotely access by law enforcing agencies. Distributed Ledger could be a very useful to handle this case with integrated with millions of devices  \cite{selim}\cite{ziacomsoc}.  
    \newline
    
    \item \textbf{\textit{Case 3}}: Let us consider a personal fitness tracker regularly recording health care dart demands to share with the individual whom it belongs, researcher and medical personnel. Besides, individual may be willing to get services accordingly from manufacturer with micro premium business relationship. The similar scenario could be though when smart homes have weather station/air monitoring IoT appliances shared with several parties.  In a case where a machine-manufacturer-practitioner-researcher network seems excessively large distributed ledger may be only smooth solution confronting challenges.  \newline
    
    \item \textbf{\textit{Case 4}}: Even blockchain cand be put in the smart electricity grids to read the total amount of energy produced by the micro-generator such as solar farm or wind turbines and also can record the dissipation time period based which net supplier payment would be issuing. Here distributed ledger can provide immutable records auditable from either ends and smart contract can make sure efficient payment process according to the stipulated rate.  \newline
    
\end{itemize}
 

\section{Conclusion}

Applying Blockchain towards efficient and scalable solution of smart and sensor based appliances of Internet of things is an emerging research area that have been rapidly evolving with an immense potentials and brand-new challenges \cite{atiur}. There are huge skepticism with on how efficiently it could be incorporated with usual IoT appliances by ensuring maximum throughput and anonymity. The effort so far made throughout this article can help novice research and developers in this arena by introducing different extant blockchain platforms and some concerning challenges in general before adopting with IoT devices. Lastly, it has brought some relevant use cases that could be considered while working on IoT leading blockchain. The paper brings a critical analysis on how Blockchain platform such as Bitcoin, Ethereum, and IOTA can be adopted for IoT applications. It concludes that all of those have immense potentials to be used as a development platform purposing to enable effective and real time deployment of smart devices on the distributed network. IOTA has been sought to be more efficient to solve transaction-latency and mining reward issue, which could be promising in different relevant use cases  to save cost by bringing throughput and efficiency. As private, public and protected blockchain have respective merits and trade-offs in different cases, therefore further research could be made to specify the exact gaps in between \cite{rebecaXun}. If the challenges and issues aroused could be minimized, it could be a great mean and benefactor of the future technology driven world.  

%
%
%
%

\bibliography{sample}
\bibliographystyle{ieeetr}
\end{document}